\documentclass[
 reprint,
 amsmath,
 amssymb,
 aps, onecolumn, nofootinbib
]{revtex4-2}

\usepackage{graphicx}
\usepackage{dcolumn}
\usepackage{bm}
\usepackage{color}
\usepackage{float}

\DeclareMathAlphabet{\pazocal}{OMS}{zplm}{m}{n}

\def\imo{i}
\def\re#1{{\rm Re}(#1)}

\def\K{{\cal K}}

\begin{document}

\title{Long-lived quasinormal modes and overtones' behavior of the holonomy corrected black holes}

\author{S. V. Bolokhov}
\email{bolokhov-sv@rudn.ru}
\affiliation{Peoples' Friendship University of Russia (RUDN University), 6 Miklukho-Maklaya Street, Moscow, 117198, Russia}


\begin{abstract}
Recently massless test scalar field perturbations of the holonomy corrected black holes [Z. Moreira et. al. Phys. Rev. D 107 (2023) 10, 104016] were studied in order to estimate  quantum corrections to the quasinormal spectrum of a black hole. Here we study both the fundamental mode and overtones of scalar, electromagnetic and Dirac fields with the help of the Leaver method and higher order WKB formula with Pad\'e approximants. We observe that the overtones depend on the geometry near the event horizon, while the fundamental mode is localized near the peak of the potential barrier, what agrees with previous studies. We showed that unlike a massless field, the massive one possesses arbitrarily long lived modes. We also obtain the analytical eikonal formula for quasinormal modes and its extension beyond eikonal approximation as a series in powers of $1/\ell$, where $\ell$ is the multipole number.
\end{abstract}

\maketitle


\section{\label{sec:Introduction}Introduction}

Quasinormal modes of black holes are crucial phenomena in the field of astrophysics and general relativity. These modes represent the characteristic oscillations that arise when perturbations are introduced to a black hole \cite{Konoplya:2011qq,Berti:2009kk,Kokkotas:1999bd,Nollert:1999ji}. They are essential for comprehending the fundamental properties of black holes, including their mass, spin, and charge, as they manifest through the distinct frequencies and damping times of these modes.

The study of quasinormal modes is imperative for scientists as it enables a deeper examination of black hole structure and offers a means to test the predictions of Einstein's theory of general relativity in the strong field regime. These modes serve as indispensable tools for gravitational wave astronomy, allowing for the detection and analysis of black hole mergers through the gravitational waves they emit \cite{LIGOScientific:2016aoc,LIGOScientific:2016sjg}. Moreover, quasinormal modes are instrumental in investigating the possibility of exotic black hole solutions, including those that arise from modifications to general relativity or theories involving extra dimensions (see, for instance,  \cite{Kanti:2006ua,Abdalla:2006qj,Seahra:2004fg,Zinhailo:2018ska,Konoplya:2020jgt,Pierini:2021jxd,Manfredi:2017xcv,Konoplya:2019hml} and references therein).

Furthermore, quasinormal modes provide a unique avenue to probe quantum gravity effects near black holes. In summary, quasinormal modes of black holes are pivotal in advancing our understanding of the fundamental nature of black holes, while also serving as a means to test the boundaries of quantum gravity.

A particular black hole model we are interested here is suggested in \cite{Alonso-Bardaji:2022ear,Alonso-Bardaji:2021yls} and motivated by the holonomy corrections owing to quantum gravity. Various properties of these black holes have been recently studied in \cite{Soares:2023uup,Junior:2023xgl}, while the first study of the quasinormal spectrum of the massless scalar and electromagnetic fields was done in \cite{Fu:2023drp} with the WKB and time-domain integration methods with the emphasis to the eikonal limit. 
A more detailed and accurate study of a test massless scalar field with the help of the lower order WKB method and Frobenius approach was done in \cite{Moreira:2023cxy}. 

Here we make the next step and study in details quasinormal frequencies of scalar (both massless and massive), electromagnetic and Dirac fields in the vicinity of such black holes with the help of the accurate convergent Frobenius method and check them with the higher order WKB method with Pad\'e approximants. We will pay special attention to the two aspects of the quasinormal spectrum: arbitrary long lived modes which take place for massive fields \cite{Konoplya:2005hr,Ohashi:2004wr} and overtones behavior. The overtones are important, because they are necessary to reproduce the ringdown signal not only at the late stage, but at the beginning of the ringing as well \cite{Giesler:2019uxc,Oshita:2021iyn}. The massive term appears as an effective term in various higher dimensional scenarios \cite{Seahra:2004fg,Ishihara:2008re} and when introducing magnetic fields around black holes \cite{Konoplya:2007yy,Konoplya:2008hj}. Among other observations, we will show that the first few overtones deviate from their Schwarzschild limit at a smaller rate than the fundamental mode, while at sufficient high $n$ the situation is opposite: the overtones differs from the Schwarzschild ones more than the fundamental mode $n=0$. We will show that this behavior is supported by the claim made in \cite{Konoplya:2022pbc}, that the overtones are determined mainly by the geometry near the event horizon, while the fundamental mode is fixed by the scattering properties near the peak of the effective potential, which is at some distance from the black hole.     

The paper is organized as follows. Sec. II is devoted to the general setup related to the background metric, wave equations and boundary conditions. Sec. III is a brief summary on the methods used for finding quasinormal modes: Frobenius method and higher order WKB approach with Pad\'e approximants. Sec. IV relates the results on numerical calculations of quasinormal modes, while Sec. V is about analytical formulas for quasinormal modes in the eikonal limit and beyond it.   

\section{General setup}

Given a general spherically-symmetric BH with the metric
\begin{equation}
ds^2=-A(r)dt^2+B(r)dr^2+r^2(d\theta^2+\sin^2\theta d\phi^2),
\end{equation}
we have the following general expressions for effective potentials of scalar, electromagnetic and Dirac fields respectively: 
\begin{eqnarray}
V_{\rm sc}(r)&=&A(r)\frac{\ell(\ell+1)}{r^2}+\frac{1}{2r}\frac{d}{dr}\frac{A(r)}{B(r)}+A(r)\mu^2,\\
V_{\rm em}(r)&=&A(r)\frac{\ell(\ell+1)}{r^2},\\
V_{\rm D\pm}(r)&=&\frac{A(r)k^2_{\pm}}{r^2}\mp\frac{A(r)k_{\pm}}{r^2\sqrt{B(r)}}\pm\frac{k_{\pm} A'(r)}{2r\sqrt{B(r)}},
\end{eqnarray}
where the multipole numbers $\ell=0,1,2,3...$ for the scalar field, $\ell=1,2,3...$ for electromagnetic field, and $k_{\pm}=1,2,3,...$ for the Dirac field.  The quantity $\mu$ is the mass of the scalar field. Here the coordinate $r$ is supposed to be a function of the ``tortoise'' coordinate $r_*$ via the relation
\begin{equation}
dr_*=\frac{dr}{f(r)}, \quad f(r)\equiv\sqrt{A(r)/B(r)}.
\end{equation}
The general master equation for calculating quasinormal modes (QNM) $\omega$ has the form
\begin{equation}
\frac{d^2\Psi}{dr_*^2}+(\omega^2-V(r))\Psi=0
\end{equation}
with the standard boundary conditions for the QNM problem:
which are requirement of purely incoming waves at the event horizon and purely outgoing ones at infinity.

In our case the metric functions have the form \cite{Alonso-Bardaji:2022ear,Alonso-Bardaji:2021yls}
\begin{equation}
A(r)=1-2M/r, \quad B(r)=\frac{1}{(1-r_0/r)A(r)}.
\end{equation}
Choosing $M=1/2$ we have the normalized horizon radius $r_h=1$.

\begin{figure}[H]
\centering
\resizebox{\linewidth}{!}{\includegraphics{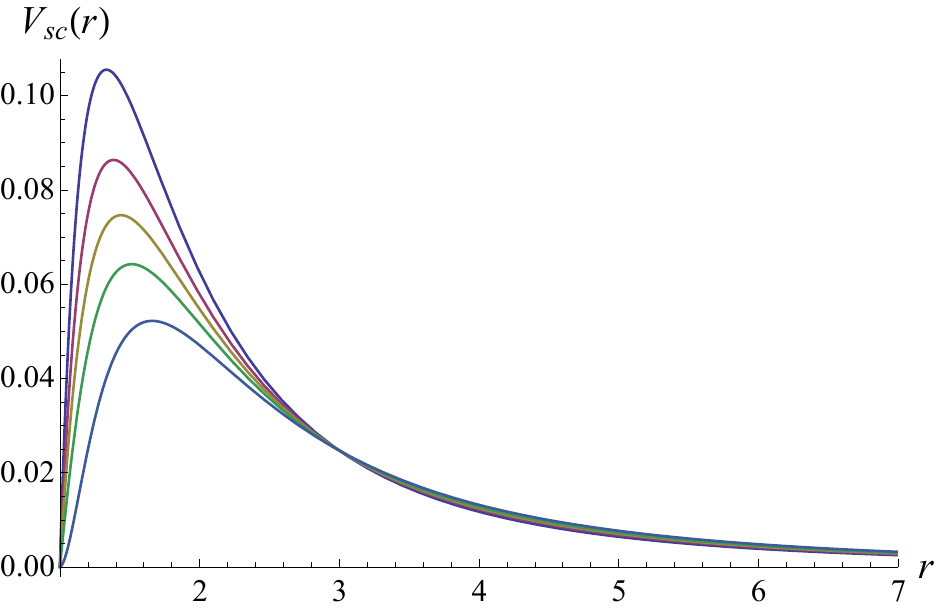}\includegraphics{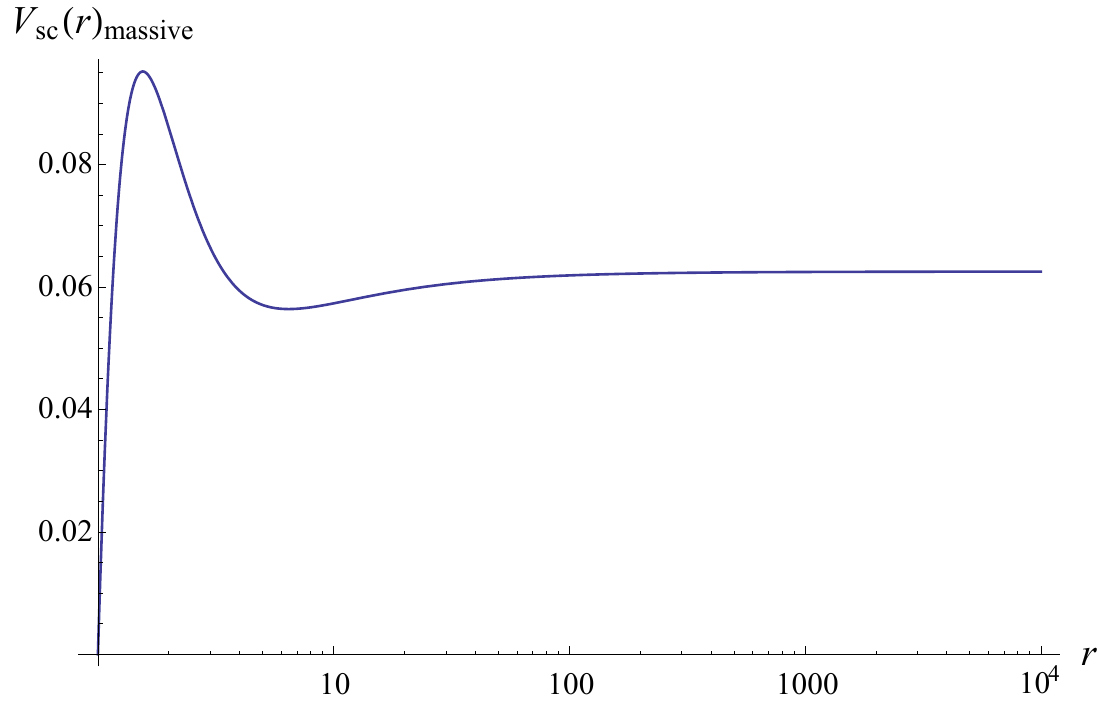}}
\caption{\textit{Left panel}: Effective potentials of a massless scalar field for $\ell=0$ and $r_0=0, 0.3, 0.5, 0.7, 0.99$ (from top to bottom), $r_h=1$. \textit{Right panel}: Effective potential of a massive scalar field for $\ell=0$, $r_0=0.5$, $\mu=0.25$; $r_h=1$.}
\label{figVsc}
\end{figure}

\begin{figure}[H]
\centering
\resizebox{\linewidth}{!}{\includegraphics{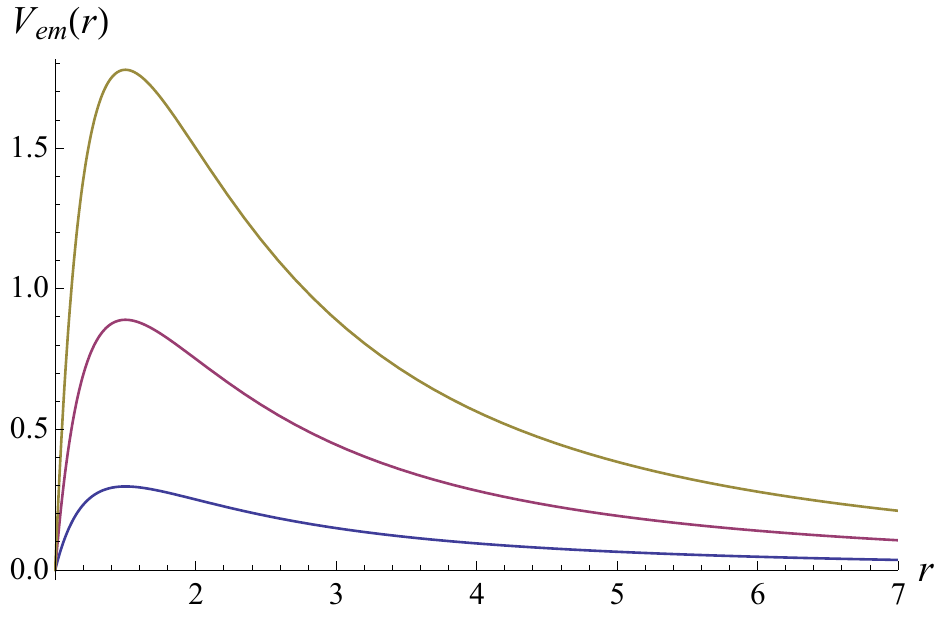}\includegraphics{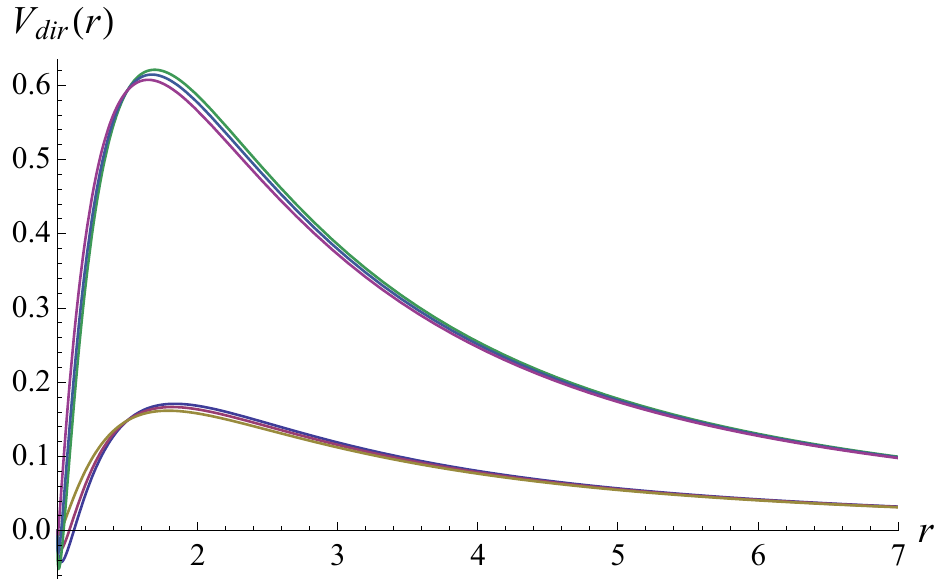}}
\caption{\textit{Left panel}: Effective potentials of electromagnetic field for $\ell=1,2,3$ (from bottom to top), $r_h=1$. \textit{Right panel}: Effective potential $V_{D-}(r)$ of the Dirac field for $k=1$ (bottom bundle of curves, $r_0=0.1, 0.5, 0.9$ from top to bottom), and $k=2$ (top bundle of curves, $r_0=0.1, 0.5, 0.9$ from top to bottom), $r_h=1$.}
\end{figure}

The effective potentials for massless fields have the form of positive definite potential barriers with a single maximum (see figs. 1-2), except one of the chiralities of the Dirac case, for which a negative gap near the event horizon takes places, as shown in fig. 3. However, the potential with the gap is iso-spectral with the other chirality case, so that it is sufficient to consider only one of the chiralities. The massive term adds another maximum far from the black hole as shown in Fig.~\ref{figVsc}.  

\section{WKB and Frobenius methods for QNMs}

Both used here methods are used and discussed in a great number of works. Therefore, here we will briefly summarize these methods, while further details can be found in   \cite{Konoplya:2011qq,Konoplya:2019hlu}.

\subsection{Frobenius method}

To acquire precise values of quasinormal modes, including overtones with $\ell < n$, we will employ the Frobenius method, also known as the Leaver method \cite{Leaver:1985ax}. The main point of the method is that the solution to the second order differential equation can be expanded into Frobenius series around the regular singular point. It is required that the coefficients in the wave equation have polynomial form, so that it can be applied to the scalar and electromagnetic fields, but not to the Dirac field in the representation we used here. The master wave-like differential equation has a regular singular point at the event horizon $r=r_0$, and an irregular one  at $r=\infty$. Therefore,  similar to the approach in \cite{Konoplya:2004uk}, we introduce a novel wave function:
\begin{equation}\label{reg}
\Psi(r)= P (r, \omega) \left(1-\frac{r_0}{r}\right)^{-\imo\omega/F'(r_0)}y(r),
\end{equation}
where we choose $P(r, \omega)$ in such a way that $y(r)$ is regular in the interval $r_0\leq r<\infty$, ensuring that $\Psi(r)$ complies with the quasinormal mode boundary conditions, which require purely incoming waves at the event horizon and purely outgoing waves at infinity. Subsequently, we express $y(r)$ in terms of a Frobenius series as follows:
\begin{equation}\label{Frobenius}
y(r)=\sum_{k=0}^{\infty}a_k\left(1-\frac{r_0}{r}\right)^k.
\end{equation}
In order to find the coefficients $a_k$ quickly in the above expansion we apply Gaussian elimination to the recurrence relation governing the expansion coefficients. This way the problem is simplified to a solution of an algebraic equation depending on $\omega$. We will also use the Nollert improvement \cite{Nollert:1993zz} in  the general n-term recurrence relation form \cite{Zhidenko:2006rs} in order to speed up the convergence of the procedure.
When singular points of the wave-like equation are located inside  the unit circle $|x| < 1$, we employ integration through a series of positive real midpoints as it was suggested in  \cite{Rostworowski:2006bp}.

\subsection{WKB approach}

The semi-analytic WKB approach was applied for the first time by  \cite{Schutz:1985km} to finding of quasinormal frequencies of asymptotically flat black holes. Then, the first order formula of \cite{Schutz:1985km} was extended to higher orders in subsequent papers \cite{Iyer:1986np, Konoplya:2003ii, Matyjasek:2017psv} and seemingly improved by using the Pad\'e approximants \cite{Matyjasek:2017psv, Hatsuda:2019eoj}.

The general form of the WKB formula can be written symbolically as an expansion around the first order, eikonal, limit \cite{Konoplya:2019hlu}:
\begin{eqnarray}
\omega^2&=&V_0+A_2(\K^2)+A_4(\K^2)+A_6(\K^2)+\ldots \\\nonumber
&-& \imo \K\sqrt{-2V_2}\left(1+A_3(\K^2)+A_5(\K^2)+A_7(\K^2)+\ldots\right),
\end{eqnarray}
where $\K=n+1/2$ is  half-integer. The corrections $A_k(\K^2)$, of order $k$, in the eikonal formula are polynomials of $\K^2$ with rational coefficients.  The explicit form of these correction terms at various WKB orders is written in the above works \cite{Iyer:1986np, Konoplya:2003ii, Matyjasek:2017psv} 

These corrections are contingent upon the values of higher-order derivatives of the effective potential $V(r)$ at its maximum, denoted as $V_j,\ (j=0,1,2,...$). In order to enhance the precision of the WKB formula, we employed the aforementioned approach presented by Matyjasek and Opala \cite{Matyjasek:2017psv}, which utilizes Pad\'e approximants. Specifically, we utilized the sixth-order WKB technique with $\tilde{m}=5$, where $\tilde{m}$ signifies the corresponding polynomial degrees within the expressions for Pad\'e approximants~\cite{Matyjasek:2017psv, Konoplya:2019hlu}. This selection provides the most accurate results in the Schwarzschild limit. Usually when the WKB frequency does not change much when one changes the WKB order or the Pad\'e splitting, say, $5$th or $6$th orders with $\tilde{m}=4$ or $5$ lead to almost the same frequencies, then this WKB result is usually in a very good agreement with the precise values (see, for example, \cite{Kodama:2009bf,Bolokhov:2023dxq,Bolokhov:2023ruj}). Thus, we use this "plateau of relative convergence" as an indicator of trusting to the WKB results.

However it's crucial to emphasize that the convergence of the WKB series is purely asymptotic and the WKB method doesn't guarantee increased precision at each order. Additionally, it does not insure the precise calculation of $\ell < n$ overtones, though in some cases first few such overtones could be found with reasonable accuracy. Therefore, in this context, we will employ the WKB method mainly as an additional check for the Leaver method when dealing with the fundamental mode $n=0$ and also in order to find  an initial guess for the frequency which is can be calculated precisely with the Leaver method.

\section{Quasinormal modes}

\subsection{Scalar field case}

Quasinormal modes for the scalar field will be considered here for both massless and massive case. As can be seen from the tables I-III, the frequencies obtained by the 6th order WKB with the Pad\'e approximants are quite close to the accurate Frobenius data, keeping the error within a small fraction of one percent, except the quasi-extreme case, for which the relative error in the damping rate achieves $1 \%$ .   Results obtained by the usual 6th (here) and 3d (in \cite{Moreira:2023cxy}) orders WKB method are much less accurate,  while the time-domain integration does not allow one to extract frequency with sufficient accuracy at $\ell=0$ perturbations \cite{Fu:2023drp}. We can see that both the real oscillation frequency and the damping rate decrease as the quantum correction parameter is increased. The damping rate is much more sensitive to the change of $r_0$ than the $\re\omega$. 

\begin{figure}[H]
\resizebox{\linewidth}{!}{\includegraphics{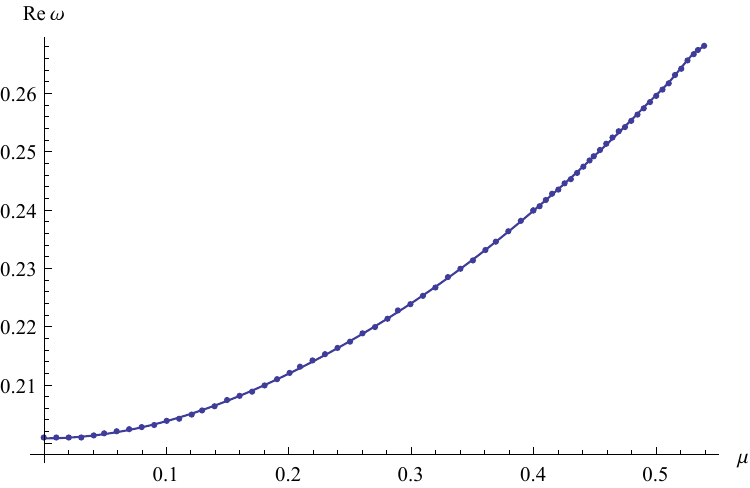}\includegraphics{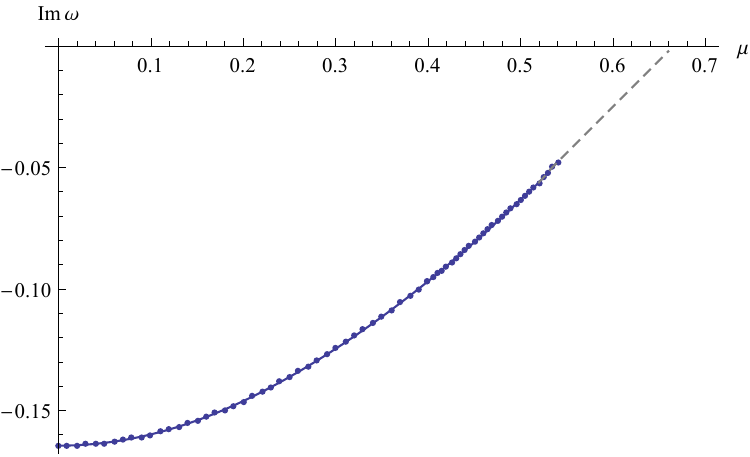}}
\caption{The real and imaginary parts of the fundamental mode $n=\ell=0$ as functions of the mass $\mu$, obtained by the Frobenius method for a massive scalar field.}
\label{fig4}
\end{figure}

When the mass term $\mu$ is tuned on, the damping rate monotonically decreases, so that there is a clear indication of the arbitrarily long lived modes, called quasi-resonances (see Fig.~\ref{fig4}). The existence  of  such modes is not guaranteed for every background. For example,  they are not allowed for the Schwarzschild-de Sitter spacetime \cite{Konoplya:2004wg}.

An interesting aspect is related to the behaviour of overtones shown on table IV: while the fundamental mode deviate from its Schwarzschild limit seemingly, by $1.6 \%$, the first few overtones change at a much smaller, but increasing with $n$ rate, so that at higher $n$ the situation becomes the opposite, i.e. the overtones deviate from their classical limits much more than the fundamental mode.  This behaviour agrees with observation made in \cite{Konoplya:2022pbc,Konoplya:2022hll}, which says that while the fundamental mode is determined by the geometry near the peak of the effective potential, the overtones are determined by the geometry in the near-horizon  zone.   Indeed, the effective potentials (see figs. 1-3) strongly deviate from the Schwarzschild ones near the maximum, but only slightly near the horizon. Therefore, the fundamental mode deviate a lot, while the first few  overtones deviate at smaller rate, though, since the phenomenon called "outburst of overtones" \cite{Konoplya:2022pbc} takes place even at small deformation in the near-horizon zone, the rate of deviation increases with $n$, so that at sufficiently high $n$ the frequencies differ much more from the Schwarzschild ones than the fundamental mode.

\begin{table}[H]
\centering
\begin{tabular}{|c|c|c|c|c|c|}
\hline
 $r_0$ & 6WKB & 6WKB(Pade6) & Frobenuis & $\delta_\text{Re}, \%$ & $\delta_\text{Im}, \%$ \\
\hline
0. & 0.220934-0.201633 i & 0.221357-0.208847 i & 0.220910-0.209791 i & 0.2 & 0.5 \\
 0.1 & 0.217552-0.193567 i & 0.217915-0.200040 i & 0.217504-0.200937 i & 0.2 & 0.4 \\
 0.2 & 0.213934-0.185853 i & 0.214249-0.191100 i & 0.213874-0.191970 i & 0.2 & 0.5 \\
 0.3 & 0.210448-0.178204 i & 0.210136-0.182227 i & 0.209951-0.182897 i & 0.09 & 0.4 \\
 0.4 & 0.208655-0.167544 i & 0.205219-0.173611 i & 0.205638-0.173735 i & 0.2 & 0.07 \\
 0.5 & 0.208603-0.151038 i & 0.200411-0.164719 i & 0.200799-0.164527 i & 0.2 & 0.1 \\
 0.6 & 0.203865-0.134380 i & 0.196607-0.155284 i & 0.195235-0.155375 i & 0.7 & 0.06 \\
 0.7 & 0.190388-0.126561 i & 0.189317-0.142897 i & 0.188651-0.146532 i & 0.4 & 2. \\
 0.8 & 0.177940-0.124934 i & 0.179155-0.136701 i & 0.180457-0.138822 i & 0.7 & 2. \\
 0.9 & 0.171056-0.122278 i & 0.172101-0.132516 i & 0.172187-0.133779 i & 0.05 & 0.9 \\
 0.99 & 0.166036-0.118817 i & 0.166942-0.128819 i & 0.166989-0.130381 i & 0.03 & 1. \\
\hline
\end{tabular}
\caption{WKB QNMs of the massless scalar field for $n=0$, $\ell=0$ and various $r_0$. The relative errors between the Frobenius and WKB6(Pade) data are denoted by $\delta_\text{Re}$ and $\delta_\text{Im}$.}
\end{table}

\begin{table}[H]
\centering
\begin{tabular}{|c|c|c|c|c|c|}
\hline
 $r_0$ & 6WKB & 6WKB(Pade6) & Frobenuis & $\delta_\text{Re}, \%$ & $\delta_\text{Im}, \%$ \\
\hline
 0. & 0.585819-0.195523 i & 0.585864-0.195320 i & 0.585872-0.195320 i & 0.001 & 0.0002 \\
 0.1 & 0.584762-0.188491 i & 0.584783-0.188359 i & 0.584781-0.188354 i & 0.0003 & 0.003 \\
 0.2 & 0.583633-0.181290 i & 0.583644-0.181190 i & 0.583630-0.181193 i & 0.002 & 0.001 \\
 0.3 & 0.582424-0.173882 i & 0.582414-0.173797 i & 0.582405-0.173811 i & 0.002 & 0.008 \\
 0.4 & 0.581117-0.166233 i & 0.581089-0.166167 i & 0.581087-0.166180 i & 0.0003 & 0.008 \\
 0.5 & 0.579685-0.158305 i & 0.579648-0.158255 i & 0.579649-0.158265 i & 0.0002 & 0.006 \\
 0.6 & 0.578086-0.150056 i & 0.578050-0.150017 i & 0.578052-0.150023 i & 0.0003 & 0.004 \\
 0.7 & 0.576262-0.141436 i & 0.576234-0.141402 i & 0.576235-0.141405 i & 0.0002 & 0.002 \\
 0.8 & 0.574121-0.132394 i & 0.574107-0.132354 i & 0.574273-0.131821 i & 0.03 & 0.4 \\
 0.9 & 0.571523-0.122887 i & 0.571527-0.122835 i & 0.571522-0.122834 i & 0.0008 & 0.0006 \\
 0.99 & 0.568621-0.113936 i & 0.568634-0.113865 i & 0.568635-0.113867 i & 0.0001 & 0.002 \\
\hline
\end{tabular}
\caption{WKB QNMs of  the massless scalar field for $n=0$, $\ell=1$ and various $r_0$. The relative errors between the Frobenius and WKB6(Pade) data are denoted by $\delta_\text{Re}$ and $\delta_\text{Im}$.}
\end{table}

\begin{table}[H]
\centering
\begin{tabular}{|c|c|c|c|c|c|}
\hline
 $r_0$ & 6WKB & 6WKB(Pade6) & Frobenuis & $\delta_\text{Re}, \%$ & $\delta_\text{Im}, \%$ \\
\hline
 0. & 0.967284-0.193532 i & 0.967287-0.193518 i & 0.967288-0.193518 i & 0.0001 & 0.0001 \\
  0.1 & 0.966641-0.186843 i & 0.966643-0.186830 i & 0.966644-0.186830 i & 0.00007 & 0.0001 \\
  0.2 & 0.965963-0.179931 i & 0.965965-0.179920 i & 0.965965-0.179920 i & 0.00004 & 0.0001 \\
  0.3 & 0.965243-0.172771 i & 0.965244-0.172761 i & 0.965244-0.172761 i & 0.00002 & 0.0001 \\
  0.4 & 0.964467-0.165328 i & 0.964467-0.165320 i & 0.964467-0.165320 i & $7.\times 10^{-6}$ & 0.0001 \\
  0.5 & 0.963619-0.157565 i & 0.963619-0.157558 i & 0.963619-0.157558 i & 0.00002 & 0.00008 \\
  0.6 & 0.962675-0.149431 i & 0.962675-0.149425 i & 0.962675-0.149425 i & 0.00003 & 0.00004 \\
  0.7 & 0.961598-0.140865 i & 0.961598-0.140860 i & 0.961598-0.140860 i & 0.00003 & 0.00001 \\
  0.8 & 0.960332-0.131789 i & 0.960332-0.131785 i & 0.963197-0.130726 i & 0.3 & 0.8 \\
  0.9 & 0.958783-0.122110 i & 0.958784-0.122106 i & 0.958783-0.122106 i & 0.00002 & 0.00007 \\
  0.99 & 0.957022-0.112793 i & 0.957022-0.112790 i & 0.957022-0.112790 i & $9.\times 10^{-6}$ & 0.00008 \\
\hline
\end{tabular}
\caption{WKB QNMs of the massless scalar field for $n=0$, $\ell=2$ and various $r_0$. The relative errors between the Frobenius and WKB6(Pade) data are denoted by $\delta_\text{Re}$ and $\delta_\text{Im}$.}
\end{table}


\begin{table}[H]
\centering
\begin{tabular}{|c|c|c|c|}
\hline
 \text{n} & \text{Holonomy corrected BH} & \text{Schwarzschild BH} & $\delta_\text{Re}, \%$ \\
\hline
0 & 0.217504-0.200937 i & 0.220910-0.209791 i & 1.6 \\
 1 & 0.172460-0.664156 i & 0.172234-0.696105 i & 0.1 \\
 2 & 0.153352-1.145927 i & 0.151484-1.202157 i & 1.2 \\
 3 & 0.143758-1.626790 i & 0.140820-1.707355 i & 2.0 \\
 4 & 0.137896-2.106328 i & 0.134149-2.211264 i & 2.7 \\
 5 & 0.133881-2.584943 i & 0.129483-2.714279 i & 3.3 \\
 6 & 0.130922-3.062923 i & 0.125988-3.216683 i & 3.8 \\
 7 & 0.128627-3.540451 i & 0.123242-3.718654 i & 4.2 \\
 8 & 0.126777-4.017645 i & 0.121013-4.220307 i & 4.5 \\
 9 & 0.125240-4.494585 i & 0.119154-4.721719 i & 4.9 \\
\hline
\end{tabular}
\caption{QNMs of massless scalar field for the holonomy corrected BH ($r_0=0.1$) and the Schwarzschild BH ($r_0=0$) for $\ell=0$ and various overtones $n$. The relative difference for ${\rm Re}\ \omega$ between these two cases is denoted by $\delta_\text{Re}$.}
\end{table}

\subsection{Electromagnetic field case}
For electromagnetic perturbations there are some distinctions from the scalar case.
First of all, $\re\omega$ monotonically increases as $r_0$ is increased, unlike the scalar case, for which it monotonically decreases. As  $\ell =1$ is the minimal value of the multipole number now, the accuracy of the WKB method with Pad\'e approximants is really great now as the comparison with the accurate Frobenius data on tables V-VI shows. An important difference with the scalar case is connected to the overtones behavior and can be easily explained. The overtones  deviate from the Schwarzschild values at a higher rate than the fundamental mode already starting from the first overtone, though this deviation develops slowly, because for a chosen value of the parameters the difference in the geometry near the event horizon is quite small. The effective potential as function of $r$is not affected by $r_{0}$ at all and the difference in geometries of the Schwarzschild and holonomy-corrected black hole is encoded in the tortoise coordinate only. For larger values of $r_{0}$ the difference of the overtones values grows much quicker with $n$.





\begin{table}[H]
\centering
\begin{tabular}{|c|c|c|c|c|c|}
\hline
 $r_0$ & 6WKB & 6WKB(Pade6) & Frobenuis & $\delta_\text{Re}, \%$ & $\delta_\text{Im}, \%$ \\
\hline
 0. & 0.496383-0.185274 i & 0.496509-0.184961 i & 0.496527-0.184975 i & 0.004 & 0.008 \\
  0.1 & 0.499896-0.179638 i & 0.500019-0.179280 i & 0.500035-0.179291 i & 0.003 & 0.007 \\
  0.2 & 0.503368-0.173640 i & 0.503482-0.173268 i & 0.503496-0.173276 i & 0.003 & 0.005 \\
  0.3 & 0.506777-0.167250 i & 0.506879-0.166889 i & 0.506890-0.166895 i & 0.002 & 0.004 \\
  0.4 & 0.510100-0.160433 i & 0.510186-0.160101 i & 0.510195-0.160105 i & 0.002 & 0.003 \\
  0.5 & 0.513301-0.153142 i & 0.513371-0.152852 i & 0.513377-0.152855 i & 0.001 & 0.002 \\
  0.6 & 0.516331-0.145320 i & 0.516383-0.145079 i & 0.516387-0.145079 i & 0.0007 & 0.0003 \\
  0.7 & 0.519115-0.136893 i & 0.519150-0.136703 i & 0.519151-0.136701 i & 0.0003 & 0.002 \\
  0.8 & 0.521532-0.127774 i & 0.521552-0.127634 i & 0.521732-0.126988 i & 0.03 & 0.5 \\
  0.9 & 0.523386-0.117876 i & 0.523396-0.117779 i & 0.523384-0.117771 i & 0.002 & 0.008 \\
  0.99 & 0.524316-0.108278 i & 0.524322-0.108206 i & 0.524295-0.108205 i & 0.005 & 0.0008 \\
\hline
\end{tabular}
\caption{WKB QNMs of electromagnetic field for $n=0$, $\ell=1$ and various $r_0$. The relative errors between the Frobenius and WKB6(Pade) data are denoted by $\delta_\text{Re}$ and $\delta_\text{Im}$.}
\end{table}

\begin{table}[H]
\centering
\begin{tabular}{|c|c|c|c|c|c|}
\hline
 $r_0$ & 6WKB & 6WKB(Pade6) & Frobenuis & $\delta_\text{Re}, \%$ & $\delta_\text{Im}, \%$ \\
\hline
0. & 0.915187-0.190022 i & 0.915190-0.190009 i & 0.915191-0.190009 i & 0.0001 & 0.0001 \\
 0.1 & 0.917186-0.183746 i & 0.917189-0.183731 i & 0.917190-0.183731 i & 0.0001 & 0.0001 \\
 0.2 & 0.919149-0.177203 i & 0.919152-0.177188 i & 0.919152-0.177188 i & 0.00008 & 0.0001 \\
 0.3 & 0.921067-0.170365 i & 0.921070-0.170351 i & 0.921070-0.170350 i & 0.00006 & 0.00009 \\
 0.4 & 0.922928-0.163195 i & 0.922930-0.163182 i & 0.922931-0.163182 i & 0.00005 & 0.00008 \\
 0.5 & 0.924714-0.155649 i & 0.924716-0.155637 i & 0.924716-0.155637 i & 0.00003 & 0.00007 \\
 0.6 & 0.926399-0.147669 i & 0.926401-0.147660 i & 0.926401-0.147659 i & 0.00001 & 0.00007 \\
 0.7 & 0.927945-0.139187 i & 0.927946-0.139179 i & 0.927946-0.139179 i & $8.\times 10^{-6}$ & 0.00007 \\
 0.8 & 0.929290-0.130111 i & 0.929291-0.130105 i & 0.930496-0.128371 i & 0.1 & 1. \\
 0.9 & 0.930333-0.120328 i & 0.930333-0.120325 i & 0.930332-0.120325 i & 0.00009 & 0.00009 \\
 0.99 & 0.930868-0.110811 i & 0.930868-0.110808 i & 0.930867-0.110809 i & 0.0001 & 0.0007 \\
\hline
\end{tabular}
\caption{WKB QNMs of electromagnetic field for $n=0$, $\ell=2$ and various $r_0$. The relative errors between the Frobenius and WKB6(Pade) data are denoted by $\delta_\text{Re}$ and $\delta_\text{Im}$.}
\end{table}


\begin{table}[H]
\centering
\begin{tabular}{|c|c|c|c|}
\hline
 \text{n} & \text{Holonomy corrected BH} & \text{Schwarzschild BH} & $\delta_\text{Re}, \%$ \\
\hline
 0 & 0.500035-0.179291 i & 0.496527-0.184975 i & 0.7 \\
 1 & 0.439399-0.566668 i & 0.429031-0.587335 i & 2.4 \\
 2 & 0.366907-1.008246 i & 0.349547-1.050375 i & 4.7 \\
 3 & 0.313786-1.478207 i & 0.292353-1.543818 i & 6.8 \\
 4 & 0.277119-1.955922 i & 0.253108-2.045101 i & 8.7 \\
 5 & 0.250381-2.435159 i & 0.224506-2.547851 i & 10.3 \\
 6 & 0.229739-2.914343 i & 0.202429-3.050533 i & 11.9 \\
 7 & 0.213084-3.393104 i & 0.184647-3.552798 i & 13.3 \\
 8 & 0.199192-3.871396 i & 0.169870-4.054612 i & 14.7 \\
 9 & 0.187310-4.349255 i & 0.157299-4.556018 i & 16.0 \\
\hline
\end{tabular}
\caption{QNMs of electromagnetic field for the holonomy corrected BH ($r_0=0.1$) and the Schwarzschild BH ($r_0=0$) for $\ell=1$ and various overtones $n$. The relative difference for ${\rm Re}\ \omega$ between these two cases is denoted by $\delta_\text{Re}$.}
\end{table}

\subsection{Dirac field case}

For the Dirac field the effective potentials does not have a polynomial form, so that we will find quasinormal modes with the 6th order WKB formula with and without Pad\'e approximants. As for scalar and vector fields the 6th order formula with Pad\'e approximants provided very  good concordance with the precise values, we believe that the same must be for the Dirac case for which the main condition of the applicability of the WKB method $\ell>n$ is fulfilled for the fundamental mode. The results of calculations is shown in tables VIII and IX. There we can see that there is non-monotonic, but smooth change of  $\re \omega$ which first grows and then decays as $r_0$ is increased. 





\begin{table}[H]
\centering
\begin{tabular}{|c|c|c|c|}
\hline
 $k$  & $r_0$ & \text{6WKB } & \text{6WKB(Pade6)} \\
\hline
 1 & 0. & 0.366162-0.194105 i & 0.365813-0.193991 i \\
 1 & 0.1 & 0.367621-0.187362 i & 0.367360-0.187224 i \\
 1 & 0.2 & 0.369030-0.180342 i & 0.368835-0.180209 i \\
 1 & 0.3 & 0.370371-0.173012 i & 0.370222-0.172915 i \\
 1 & 0.4 & 0.371617-0.165327 i & 0.371491-0.165303 i \\
 1 & 0.5 & 0.372731-0.157236 i & 0.372601-0.157325 i \\
 1 & 0.6 & 0.373653-0.148676 i & 0.373474-0.148926 i \\
 1 & 0.7 & 0.374286-0.139579 i & 0.373987-0.140027 i \\
 1 & 0.8 & 0.374469-0.129886 i & 0.373973-0.130546 i \\
 1 & 0.9 & 0.373946-0.119604 i & 0.373182-0.120470 i \\
 1 & 0.99 & 0.372613-0.109991 i & 0.371527-0.111055 i \\
\hline
\end{tabular}
\caption{WKB QNMs of the Dirac field with the potential $V_{D-}(r)$ for $n=0$, $k=1$ and various $r_0$.}
\end{table}

\begin{table}[H]
\centering
\begin{tabular}{|c|c|c|c|}
\hline
 $k$  & $r_0$ & \text{6WKB } & \text{6WKB(Pade6)} \\
\hline
 2 & 0. & 0.760077-0.192833 i & 0.760073-0.192813 i \\
 2 & 0.1 & 0.760875-0.186255 i & 0.760872-0.186238 i \\
 2 & 0.2 & 0.761632-0.179435 i & 0.761629-0.179421 i \\
 2 & 0.3 & 0.762336-0.172344 i & 0.762334-0.172332 i \\
 2 & 0.4 & 0.762973-0.164947 i & 0.762972-0.164936 i \\
 2 & 0.5 & 0.763521-0.157200 i & 0.763520-0.157189 i \\
 2 & 0.6 & 0.763948-0.149051 i & 0.763947-0.149038 i \\
 2 & 0.7 & 0.764206-0.140433 i & 0.764207-0.140417 i \\
 2 & 0.8 & 0.764222-0.131264 i & 0.764225-0.131243 i \\
 2 & 0.9 & 0.763872-0.121451 i & 0.763878-0.121422 i \\
 2 & 0.99 & 0.763078-0.111991 i & 0.763088-0.111955 i \\
\hline
\end{tabular}
\caption{WKB QNMs of the Dirac field with the potential $V_{D-}(r)$ for $n=0$, $k=2$ and various $r_0$.}
\end{table}


\section{Analytical formula for qnms in the eikonal limit and beyond}

The regime of high real oscillation frequency, i.e. of high multipole numbers $\ell$ is hardly do be observed in the near future, because in merger of two compact objects such as black holes or neutron stars usually only a couple of first gravitational multipole numbers ($\ell=2,3$) are excited considerably. Nevertheless,  there are  a number of reasons of theoretical character  to study this regime. First of all, the eikonal regime corresponds to the limit of geometrical optics and it comes as no surprise that in this regime there is a correspondence between  characteristics of null geodesics and  eikonal quasinormal modes \cite{Cardoso:2008bp}. More exactly, the real oscillation frequency is determined by the rotation frequency of the unstable circular null geodesic\footnote{It is also worth mentioning that and in the non-linear electrodynamics photons move not along the null geodesics, so that the terms "null geodesics" and "photons orbits" are not always interchangeable in the correspondence  \cite{Chen:2018vuw,Chen:2019dip,Toshmatov:2019gxg}.} and the damping rate is proportional to the Lyapunov exponent \cite{Cardoso:2008bp}.   However, as it was shown in \cite{Konoplya:2017wot,Konoplya:2022gjp}, this correspondence works for the WKB-like part of the spectrum and, therefore, is not always guaranteed.  Thus, for example, it is not fulfilled for the gravitational perturbations in some theories with higher curvature corrections  \cite{Konoplya:2019hml,Konoplya:2020bxa}.

The other reason to treat separately the eikonal regime is that it may bring qualitatively new features of the perturbations. Thus, for example, the  high $\ell$ modes may govern the instability, called "eikonal instability" \cite{Dotti:2005sq,Takahashi:2012np,Takahashi:2010gz,Konoplya:2017lhs},  so that the small $\ell$ frequencies which are stable in such a case do not have meaning anymore: once there is one unstable mode the system is unstable.

An advantage of the eikonal regime is that it is that rare case in which quasinormal modes can be found analytically\footnote{The other examples of spectral problems allowing for an exact solution of the wave equation are mostly for the 2+1 dimensional black holes, such as BTZ spacetime and its generalizations \cite{Banados:1992wn,Konoplya:2020ibi}.}, because the WKB method is exact in this limit. Here, using the general approach and {\it Mathematica} code shared in \cite{Konoplya:2023moy} in order to find the eikonal formula and some improvement of it via expansion beyond the eikonal limit in powers of $1/\ell$.

The maximum of the effective potential acquires the $1/\kappa^2$ correction,
\begin{equation}
r_{max} = 3 M+ \frac{1}{\kappa^2}\left(\frac{r_0}{4}-\frac{M}{3}\right) + {\cal O} (1/\kappa^3),
\end{equation}
where $\kappa =\ell+\frac{1}{2}$.
Then expansion in powers of $\kappa$ gives:
\begin{equation}
\omega = \frac{\kappa }{3 \sqrt{3} M}-\frac{i K \sqrt{3 M-r_0}}{9 M^{3/2}}-\frac{36 M^2 \left(60 K^2-29\right)+12 M r_0
   \left(79-156 K^2\right)+r_0^2 \left(492 K^2-191\right)}{15552 \sqrt{3} M^2 \kappa  (3 M-r_0)}+ {\cal O}(1/\kappa^2),
\end{equation}
where $K= n+\frac{1}{2}$.

For the electromagnetic case the effective potential does not gain a $r_0$ correction, so that the effect is only due to the corrected tortoise coordinate.  The terms beyond the eikonal approximation differs from that for the scalar field,
\begin{equation}
\omega = \frac{\kappa }{3 \sqrt{3} M}-\frac{i K \sqrt{3 M-r_0}}{9 M^{3/2}}-\frac{180 M^2 \left(12 K^2+23\right)-12 M r_0
   \left(156 K^2+173\right)+r_0^2 \left(492 K^2+241\right)}{15552 \sqrt{3} M^2 \kappa  (3
   M-r_0)} + {\cal O}(1/\kappa^2).
\end{equation}
For the Dirac field, the maximum of the effective potential is located at

\begin{equation}
r_{max} =3 M+\frac{\sqrt{M (3 M-{r_0})}}{2 \kappa }+\frac{{r_0}}{12 \kappa ^2}+ {\cal O}(1/\kappa^3) ,
\end{equation}

while the quasinormal frequency at the first order in $1/\kappa$ beyond the eikonal limit is

\begin{equation}
\omega =\frac{\kappa }{3 \sqrt{3} M}-\frac{i K \sqrt{3 M-r_0}}{9 M^{3/2}}-\frac{36 M^2\left(60 K^2+7\right) -12 M
   r_0\left(156 K^2+11\right) +r_0^2\left(492 K^2+25\right) }{15552  \sqrt{3} M^2 \kappa \left(3 M-r_0\right)}+{\cal O}(1/\kappa^2).
\end{equation}

In the eikonal limit the quasinormal frequencies  do not depend on the spin of the field under consideration. Comparison of the above formula in the eikonal limit with the Lyapunov exponents and rotational frequency for the null geodesics shows that the null geodesics/eikonal quasinormal modes correspondence is fulfilled in this case.

\section{Conclusions}

We have presented a detailed study of quasinormal modes of the holonomy corrected black holes for scalar, electromagnetic and Dirac fields, including the overtones behaviour and long-lived modes of a massive field. The quasinormal modes were computed here with two alternative methods (WKB with Pad\'e approximants and Leaver) with excellent agreement between them in the common range of applicability. We have shown that:
\begin{itemize}
    \item For a nonzero mass of the field, the quasinormal modes become much longer lived and there is a clear indication of existence of arbitrarily long lived modes - quasiresonances.
    \item The overtones behaviour is in concordance with observation that the higher is the overtone's number, the more it depends on  the geometry near the event horizon, while the fundamental mode depends upon the scattering near the peak of the effective potential.
    \item While the third order WKB method used in earlier studies of the holonomy corrected black hole spectrum agrees well with the Leaver method only at high multipole numbers, application of the 6th order with Pad\'e approximants provides concordance at all $\ell$.
    \item The analytical formula for quasinormal modes in the eikonal limit, and at some orders of $1/\ell$ beyond, was deduced and it may serve as a reasonable approximation for the lowest modes.
\end{itemize}

Outburst of overtones studied recently in a number of works also on various quantum corrected black holes \cite{Konoplya:2023ppx,Konoplya:2023aph} is characterized by a strong deviation already at the first few overtone numbers $n$. Here we observed a different picture of slow, but still increasing with $n$, deviation of overtones for both scalar and electromagnetic fields. In principle, the Dirac equation could be re-written using a different representation basis, so that one could reduce it to the polynomial form \cite{Cho:2003qe,Cho:2007zi}. Then, the overtones behavior could analyzed for the Dirac field as well.

\acknowledgments{The author would like to acknowledge Dr. R. A. Konoplya for careful reading of the manuscript and most useful discussions. This work was supported by RUDN University research project FSSF-2023-0003.}

\bibliographystyle{unsrt}
\bibliography{BH_holonomy}

\end{document}